\documentclass[aps,prb,10pt,twocolumn,floatfix,showpacs,superscriptaddress]{revtex4-1}
\usepackage{graphicx,epsfig,amssymb,color}

\begin{document}

\title{A theoretical analysis of small Pt particles on rutile TiO$_\mathbf{2}$(110) surfaces}

\author{Veysel \c{C}elik}
\author{Hatice \"{U}nal}
\author{Ersen Mete}
\email[Corresponding author email : ]{emete@balikesir.edu.tr}
\altaffiliation[ Also at ]{Institute of Theoretical and Applied Physics (ITAP)
Turun\c{c}, Mu\u{g}la, Turkey}
\affiliation{Department of Physics, Bal{\i}kesir University,
Bal{\i}kesir 10145, Turkey}
\author{\c{S}inasi Ellialt{\i}o\u{g}lu}
\affiliation{Department of Physics, Middle East Technical University, Ankara
06531, Turkey}

\date{\today}

\begin{abstract}
The adsorption profiles and electronic structures of Pt$_n$ ($n$\,=\,1\,--\,4)
clusters on stoichiometric, reduced and reconstructed rutile
TiO$_2$(110) surfaces were systematically studied using on site $d$--$d$
Coulomb interaction corrected hybrid density functional theory calculations.
The atomic structure of small Pt cluster adsorbates mainly depend on
the stoichiometry of the corresponding titania support. The cluster
shapes on the bulk terminated ideal surface look like their gas phase
low energy structures. However, for instance, they get significantly
distorted on the reduced surfaces with increasing oxygen vacancies.
On non-stoichiometric surfaces, Pt--Ti coordination becomes dominant in
the determination of the adsorption geometries. The electronic structure
of Pt$_n$/TiO$_2$(110) systems can not be correctly described by pure DFT
methods, particularly for non-stoichiometric cases, due to the
inappropriate treatment of the correlation for $d$ electrons. We performed
DFT+U calculations to give a reasonable description of the reconstructed
rutile (110) surface. Pt clusters induce local surface relaxations
that influence band edges of titania support, and bring a number of band-gap
states depending on the cluster size. Significant band gap narrowing
occurs upon Pt$_n$--surface interaction due to adsorbate driven states
on the bulk terminated and reduced surfaces. On the other hand, they
give rise to a band gap widening associated to partial reoxidation of the
reconstructed surface. No metallization arises even for Pt$_4$ on rutile.
\end{abstract}


\maketitle

\section{Introduction}

Transition metal oxides are important because of their wide range
of technological applications. Naturally occurring rutile polymorph
of titanium dioxide is a generic material because of its abundance,
non-toxicity, and stability under atmospheric conditions. The (110)
termination of rutile TiO$_2$ is energetically the most stable surface 
among the other low index facets.~\cite{henrich1} For these reasons, 
rutile TiO$_2$(110) structure is a prototypical material to understand 
the catalysis on more complex oxide surfaces.~\cite{fujishima, diebold1} 
Its reducibility (either by oxygen vacancy formation or alkali metal 
adsorption) raises a great interest for fundamental study of photo- 
and heterogeneous catalysis~\cite{hangfeldt,gratzel,khan}, functional 
ultrathin films~\cite{chen,finetti}, and dielectrics.~\cite{wu,griffin}
These applications rely mainly on the surface electronic properties
of titania.

\begin{figure}[htb]
\epsfig{file=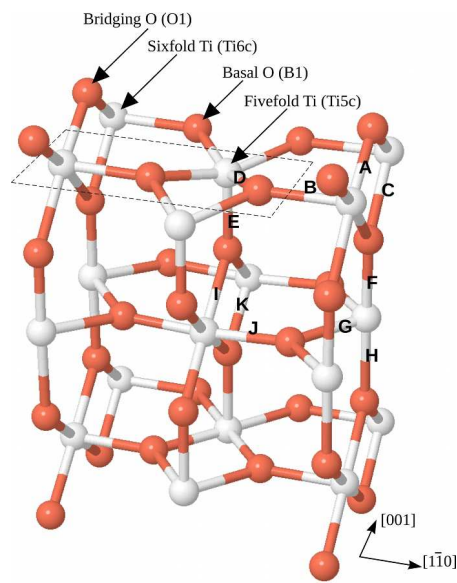,width=5.9cm}
\caption{(color online) Optimized atomic structure of bulk
terminated rutile TiO$_2$(110) surface. Actual computational
cell has 15 atomic layers also referred as 5 trilayers. The
model shows only 3 out of 5 trilayers. Black (red) and white
balls represent O and Ti atoms, respectively. Dashed rectangular
area denotes the (1$\times$1) surface unit cell. The bond labels
are indicated as capital letters that run from A to K.\label{fig1}}
\end{figure}

In TiO$_2$ bulk structure, Ti atoms are sixfold and oxygens are
threefold coordinated. Rutile (110) surface has these types of
atoms, such as basal atoms B1 and Ti6c,  which exhibit bulk like
bonding characteristics as depicted in Fig.~\ref{fig1}. The (110)
termination of the bulk lattice breaks the Ti-O bonds which lie
normal to the surface plane resulting in fivefold Ti and twofold
O atoms on the surface. Those are denoted as Ti5c and O1 in
Fig.~\ref{fig1}, respectively. The undercoordinated bridging
oxygens, O1, are exposed on the surface. They form oxygen rows
along [001] direction. The atomic disposition of this
stoichiometric long range ordered (1$\times$1) phase has been
well established by experimental methods~\cite{diebold1,see,
zzhang,novak,diebold2,charlton,asari,lindsay} and by
\emph{ab initio} calculations.~\cite{ramamoorthy1,ramamoorthy2,
lindan,kimura,harrison,vogtenhuber,reinhardt,bates,evarestov,
bandura,bredow,sano,yfzhang,kiejna,thompson,hameeuw,kowalski,
labat}

By means of ion bombardment or thermal annealing, the surface can be
reduced through oxygen removal. Its oxygen vacancy induced reducibility
enhances hetero- and photocatalytic activity by giving rise to rich
surface chemistry. In this sense, rutile (110) system becomes an
excellent model substrate that shows interesting properties of more
complex metal oxides.~\cite{henrich2,gopel,kurtz,henderson,wendt} 

Although DFT is successful in describing ground state atomic and 
electronic structure of many semiconductors, there is still a 
debate on the choice of method, pure or hybrid, employed within
the framework of DFT to improve the accuracy to estimate the
material properties. Particularly, some of DFT predicted
properties of reconstructed and reduced TiO$_2$ surfaces are
inconsistent with the experimental data.~\cite{kruger,kimmel}
This is due to inadequate description of strongly correlated
$3d$ electrons of Ti atoms.~\cite{morgan,calzado,nolan}

Since bulk termination gives rise to a sharp discontinuity in the
atomic bonds, rutile (110) is known occasionally to undergo surface
morphological reconstructions.~\cite{onishi,guo,pang,mccarty,elliot1,
elliot2,blancorey1,blancorey2,park,shibata} This phase transition
is believed to be assisted by annealing that reduces the surface
and leads to a (1$\times$2) reconstruction. The three dimensional
determination of such atomic rearrangements on reconstructed rutile
(110) surface is difficult by experimental techniques.~\cite{shibata}
Onishi and Iwasawa proposed the Ti$_2$O$_3$ added row model for the
reconstructed rutile (110) surface.~\cite{onishi} There are also some
controversial experimental results. For example, Park \textit{et al.}~\cite{park}
suggested a new model where Ti are on interstitial sites forming a
relatively more O deficient surface stoichiometry. This model recently
confirmed experimentally by Shibata \textit{et al.}~\cite{shibata}
Although it is a matter of debate, the added row model of Onishi and
Iwasawa remains to be widely assigned (1$\times$ 2) rutile (110)
reconstruction.~\cite{guo,pang,elliot1,elliot2,blancorey1}

The steps and point defects such as oxygen vacancies and (sub)surface
impurities shown in scanning tunneling microscopy (STM) images play an
important role in electronic properties of TiO$_2$ surfaces~\cite{wendt}.
Understanding of the fundamental aspects of defects and impurities on
titania surfaces are important for technological applications. In addition,
the catalytic activity of TiO$_2$ surfaces can be promoted by transition
metal (such as Au or Pt) doping~\cite{gai} or adsorption~\cite{linsebigler,iddir}.
Therefore, the interaction of surface defects with functional adsorbates
is crucial for both fundamental research and practical applications. For
example, Pillay and Hwang~\cite{pillay} studied the adsorption geometries
of small Au, Ag, and Cu clusters on rutile (110) surface. Recently,
Gong \textit{et al.}~\cite{gong} investigated the atomic structures of
small Au and Pt clusters supported on defect and regular sites of anatase
TiO$_2$(101) surface.

In this work, our primary aim is to study the adsorption sites,
geometries and resulting electronic structures of Pt$_n$ ($n$\,=\,1\,--\,4)
clusters on the stoichiometric, reduced (by bridging oxygen B1
removal) and reconstructed (added row model~\cite{onishi}) rutile
TiO$_2$(110) surfaces using \textit{ab initio} calculations. In order
to gain insight into the growth pattern of small Pt particles on these
surfaces, we chose a single Pt adsorbate as the starting point. Then,
we studied the adsorption profiles of Pt$_n$  by adding extra Pt atoms 
at different probable adsorption sites to the previously optimized 
Pt/TiO$_2$(110) system. Alternatively, we also put Pt particles, whose 
geometries were optimized in their gas phase, on probable adsorption 
sites. By comparing these two distinct approaches we decided on the 
lowest energy adsorption structures of small Pt particles on the 
surfaces considered.

\section{Computational Details}

Total energy density functional theory (DFT) calculations were
carried out using the Vienna \textit{ab-initio} simulation package
(VASP).~\cite{vasp} Nonlocal exchange--correlation energies were
treated with Perdew--Burke--Ernzerhof (PBE)~\cite{pbe} functional
based on the generalized gradient approximation (GGA). We used
projector-augmented waves (PAW) method~\cite{paw1,paw2} to
describe the ionic cores and valence electrons with an energy
cutoff value of 400 eV for the plane wave expansion.

Pure DFT methods systematically fail in describing strongly
correlated $3d$ electrons localized on Ti atoms, particularly in
the case of reduced rutile (110) surface.~\cite{morgan,calzado,nolan}
This comes from the inherent shortcoming of the DFT due to the lack
of proper self-energy cancellation between the Hartree and exchange
terms as in Hartree--Fock theory. DFT+U approach is an alternative
attempt to compensate this localization deficiency by introducing
additional Hubbard $U$ term for electron on-site repulsion. Therefore,
in order to get a sound description of Pt$_n$ adsorption on reduced
and reconstructed TiO$_2$(110) surfaces, we performed Hubbard $U$
corrected DFT calculations using Dudarev's approach.~\cite{dudarev}
We supplemented PBE exchange--correlation functional with Dudarev $U-J$
term acting on Ti 3$d$ states, that we refer as GGA+U.  This empirical
term is adjusted to give reasonable agreement with the experimental
data. Morgan and Watson fitted the $U$ parameter to 4.2 eV in order
to reproduce the experimentally observed position of oxygen vacancy
driven Ti defect state in the band gap.~\cite{morgan} In the same
manner, our tests suggested to choose the parameters $U$=4.5 eV and
$J$=0 eV to obtain a reasonable electronic description of the added
row model of Onishi and Iwasawa as well as the gap state at the
reduced surface.

Theoretical studies employing other hybrid methods such as
incorporation of the exact Fock exchange energy (EXX) with various
percentages, recently, revisited the stoichiometric rutile
TiO$_2$(110) surface.~\cite{yfzhang} Zhang \textit{et al.}
reproduced many of the experimental bulk and stoichiometric
surface properties of TiO$_2$ by adjusting the mixing of HF exchange
contribution. Further investigations are still needed to see if EXX
approach can correctly describe the defect states.

Computationally, there has been another debate about the slab
thickness to correctly predict the surface energetics as well as
the atomic structure. The surface energy is known to fluctuate
with slab thickness.~\cite{ramamoorthy2,wu,bredow,thompson,hameeuw,
kiejna,kowalski} For adsorption of small molecules on the bulk
terminated surface, Thompson \textit{et al.}~\cite{thompson}
suggested 4 layers of O--Ti$_2$O$_2$--O units (also referred as
trilayers) with the two bottom layers fixed to their bulk
positions. Kowalski \textit{et al.}~\cite{kowalski} additionally
suggested the saturation of the dangling bonds at the bottom surface
with pseudohydrogens having nuclear charges of +4/3 and +2/3,
correspondingly.

Our stoichiometric slab model consists of 15 atomic layers that
correspond to 5 layers of O--Ti$_2$O$_2$--O units. This model has
a mirror symmetry with respect to the central atomic plane. The
dashed rectangular region shown in Fig.~\ref{fig1}, represents
$p$(1$\times$1) periodicity. The corresponding surface unit cell
is composed of a central Ti$_2$O$_2$ planar arrangement with two
oxygens symmetrically cross bonded to the nearest neighbor in-plane
Ti atom from above and from below. These units are shifted by half
its length along [$\bar 1$10] between adjacent layers as seen in
Fig.~\ref{fig1} which shows three out of the five layers in our
model. The slab was separated from its images along the surface
normal by a vacuum region of $\sim$14~\AA.

For geometry optimizations, the Brillouin zone integrations were
carried out with 2$\times$2$\times$1 Monkhorst--Pack~\cite{mp} $k$-point
grid for unreconstructed (bulk terminated and reduced)
and reconstructed cells with surface periodicities $p$(4$\times$2)
and $p$(4$\times$1), respectively. The spin polarization was
found  to be negligibly small for the Pt$_n$ on stoichiometric surface
considered in this work. This is in agreement with the similar findings of
Iddir {\it et al.} for single Pt adsorbate on TiO$_2$(110)~\cite{iddir}.
We performed spin polarized GGA+U calculations for Pt$_n$/TiO$_2$ systems
where magnetization is non-negligible, in particular, for reduced
and reconstructed surface cases. A full geometry optimization was fulfilled
with residual minimization method direct inversion in the iterative subspace
(RMM-DIIS) scheme preconditioned by a few non-selfconsistent Davidson-block
iterations as implemented in the code. We required a precision of
10$^{-2}$ eV/\AA~in the residual forces in every spatial component
on all the atoms  without fixing them to their bulk positions.

We estimated the binding energies of Pt clusters by,
\[
E^{\rm b}_{{\rm Pt}_n}=E_{{\rm Pt}_n/{\rm TiO_2}}-E_{\rm TiO_2}-E_{{\rm Pt}_n}\, ,
\]
where $E_{{\rm Pt}_n/{\rm TiO_2}}$, $E_{\rm TiO_2}$ and $E_{{\rm Pt}_n}$ are the total
energies of the Pt$_n$/TiO$_2$ combined system, the bare TiO$_2$ slab and
the corresponding Pt$_n$ cluster, respectively. The average adsorption
energy per atom, $E^{\rm c}_{{\rm Pt}_n}$, can be calculated by dividing
$E^{\rm b}_{{\rm Pt}_n}$ by the cluster size $n$.

\section{Results \& Discussion}

We made extensive tests for the choice of the slab model by changing
the slab thickness, the number of fixed atomic layers within, and the
size of the vacuum spacing, provided that the (110) facet possesses
4$\times$2 cell symmetry to avoid any interaction between the periodic
images of Pt cluster adsorbates. The results suggest that no atom,
particularly near the surface, should be fixed to its bulk position
in order to avoid stress driven gap states and that the slab model
has to be at least 5 trilayers thick (15 atomic layers) to ensure
a bulk-like central part. Similarly, for instance, Kiejna
\textit{et al.}~\cite{kiejna} found that surface energetics are
very sensitive to lattice relaxations by considering the effect of
various slab model thicknesses. Other hybrid DFT approaches with
different exchange and correlation treatments are known to have a
small effect on the surface atomic configurations~\cite{yfzhang}.

\begin{table}[h]
\caption{The comparison of computational and experimental bond
lengths for the bulk terminated stoichiometric TiO$_2$(110) surface.
All measurements are in angstroms. Bond labeling follows
similar to those of Thompson {\it et al.} as shown in Fig.~\ref{fig1}.
\label{table1}}
\begin{ruledtabular}
\begin{tabular}{cccccccc}
Bond & \multicolumn{2}{c}{Experimental} & \hspace{3mm}
&\multicolumn{4}{c}{Theoretical}
\\[1mm] \cline{2-3} \cline{5-8} \\[-3mm]
& Ref.~\cite{charlton} & Ref.~\cite{lindsay} &&
Ref.~\cite{bates} & Ref.~\cite{thompson}\footnote{values for
5 layer slab model} & GGA\footnote{results of present calculation
with PBE xc functional} & GGA+U$^b$ \\[1mm] \hline
A & 1.71$\pm$0.07 & 1.85 && 1.80 & 1.84 & 1.85 & 1.88 \\
B & 2.15$\pm$0.09 & 2.15 && 2.04 & 2.04 & 2.03 & 2.03 \\
C & 1.99$\pm$0.09 & 2.08 && 2.09 & 2.11 & 2.10 & 2.09 \\
D & 1.84$\pm$0.05 & 1.90 && 1.95 & 1.92 & 1.95 & 1.95 \\
E & 1.84$\pm$0.13 & 1.79 && 1.85 & 1.83 & 1.84 & 1.90 \\
F & 1.97$\pm$0.12 & 1.90 && 1.90 & 1.89 & 1.90 & 1.95 \\
G & 1.99$\pm$0.05 & 2.00 && 1.97 & 1.98 & 1.98 & 1.97 \\
H & 2.18$\pm$0.11 & 2.11 && 2.11 & 2.13 & 2.12 & 2.12 \\
I & 2.00$\pm$0.08 & 2.01 && 2.02 & 2.02 & 2.02 & 2.01 \\
J & 1.92$\pm$0.06 & 1.92 && 1.97 & 1.96 & 1.96 & 1.96 \\
K & 1.94$\pm$0.06 & 1.89 && 1.90 & 1.91 & 1.92 & 1.95
\end{tabular}
\end{ruledtabular}
\end{table}

Indeed, a comparison of our calculated bond lengths for the
stoichiometric rutile TiO$_2$(110) surface shows a good agreement
with the available experimental~\cite{lindsay,charlton} and
theoretical~\cite{thompson,bates} results as presented in
Table~\ref{table1}. Moreover, the incorporation of supplemental
Coulomb repulsion to correct the correlation energy by inclusion
of Dudarev $U=4.5$ eV term for Ti 3$d$ electrons did not distort
the atomic positions, considerably. In addition to these structural
properties, we also calculated the binding energies (BE) of Pt to be
2.32 eV and 3.32 eV on the stoichiometric and partially reduced
(with oxygen vacancy concentration of $\theta$=1/8) surfaces,
respectively. Our values are presented in Table~\ref{table2} and are
only slightly larger than the previous theoretical results of
Iddir \textit{et al.}~\cite{iddir} This can be well addressed to the
difference in slab thicknesses and in surface areas of computation
cells. In particular, the oscillatory convergence behavior of the
surface electronic properties depending on the number of trilayers,
included in a slab calculation, was also confirmed by other
studies.~\cite{kiejna,thompson,kowalski,labat} From computational
point of view, a slab with large number of trilayers is expected
to give very well converged results for Pt clusters on 4$\times$2
cell. However, independent of slab thickness, the electronic
description of defect states such as the oxygen vacancies is
problematic with pure DFT methods.~\cite{morgan,calzado,nolan} In the
same line, our tests showed that standard DFT fails in describing
the electronic structure of the added row model of reconstructed
surface, as well, by giving surface Ti 3$d$ state in the conduction
band (CB). This study, particularly, focuses on a reasonable
electronic description of small Pt clusters on rutile (110) surface
where spin polarization is non-negligible. Therefore, our GGA+U
values for this slab model would not only give very well converged
bond lengths, as presented in Table~\ref{table1}, but also yield the
energetics reasonably accurate for a Pt cluster adsorption
system on specifically non-stoichiometric rutile (110) surfaces.

In order to study the adsorption characteristics of Pt$_n$
($n$\,=\,1\,--\,4) particles on rutile TiO$_2$(110) surfaces we
first considered these in the gas phase and performed noncollinear
spin polarized DFT calculations. A recent work stresses the necessity
for inclusion of spin--orbit interactions in a relevant calculation
to understand the physics of Pt clusters~\cite{huda}. However, the
ground state geometries being a dimer for Pt$_2$, a triangle for
Pt$_3$ and a bent (off-planar) rhombus in the case of Pt$_4$ seem
to be less affected by spin--orbit coupling.

For Pt dimer, our non-spin--orbit calculations resulted in triplet
electronic ground state with a binding energy (BE) of 1.819 eV/atom
and gave the bond length to be 2.331 {\AA}. The calculated dimer
length shows an excellent agreement with the experimental value of
2.333 {\AA} \cite{airola,taylor,grushow}, while our result for the
BE is slightly higher than the experimental value of 1.570 eV/atom.
When we included self-consistent noncollinear spin--orbit coupling
that resulted in the same magnetic ground state, the BE decreases to
1.665 eV/atom whereas the dimer bond length increases to 2.382 {\AA}.
These values are sensitive not only to the inclusion of spin--orbit
effects but also to the exchange--correlation scheme employed. For
instance, by performing GGA-PW91 calculations, Huda \textit{et al.}
\cite{huda} determined different spin multiplicities for the ground
state of Pt$_3$ cluster depending on the spin--orbit coupling. However,
our GGA-PBE calculations gave singlet ground states for Pt$_3$ having
average binding energies of 2.184 and 2.376 eV/atom with and without
LS coupling, respectively. Moreover, Huda \textit{et al.} reported
that spin--orbit interaction drives the geometry from an equilateral
to an isosceles triangle. Pt$_3$ with GGA-PBE, on the other hand, 
develops equilateral coordination with bond lengths of 2.49 {\AA}
while the inclusion of spin--orbit interaction extends all bonds
slightly and equally to 2.50 {\AA}. Our GGA-PBE binding energy
value agrees well with a previous theoretical result.~\cite{xiao}
In addition, incorporation of spin-orbit coupling leads to an
excellent agreement with an old experimental estimate of
2.18 eV/atom.~\cite{miedema}

Pt$_4$ has a quintet electronic ground state (GS) with
GGA-PBE, leading to an off-planar rhombus geometry that has a side
bond length of 2.51 {\AA} and a bending angle of 23.3$^\circ$. We
calculated atomic BEs to be 2.515 eV/atom with, and
2.686 eV/atom without the LS coupling. These values are comparable
with the previous results.~\cite{huda,yang}  However, spin--orbit
interaction included in the GGA-PBE calculation did not make the
geometry perfectly planar as reported for GGA-PW91.~\cite{huda}
It slightly extends the bond lengths to 2.52 {\AA} by decreasing
the bending angle to 13.1$^\circ$ and gives the same spin multiplicity
for the GS. Consequently, for all isolated small Pt particles
considered in this study, inclusion of LS coupling in GGA-PBE
exchange--correlation scheme does not deform the GS cluster shapes
into new geometries.

\subsection{Pt$_\mathbf{n}$ on stoichiometric rutile TiO$_\mathbf{2}$(110) surface}

Pure DFT with GGA-PBE functional gives a band gap of 1.48 eV for
the stoichiometric rutile TiO$_2$(110) surface. Supplemental Hubbard
$U$ on-site repulsion, acting on the Ti 3$d$ electrons, heals
this band gap to 2.01 eV by correcting the underestimation in the
correlation energy. Although the Dudarev $U$ parameter can be adjusted
to a larger value to get a better agreement with the experimental
value of $\sim$3 eV~\cite{henrich1}, this would significantly distort
the atomic structure. Therefore, without losing the consistency of
calculated atomic positions and bond lengths with the experimental
values (as presented in Table~\ref{table1}), we set it to $U=4.5$, in
consistency with the previous theoretical studies~\cite{calzado,morgan,nolan},
to get the Ti defect state in the band gap of the reduced surface and
to obtain a reasonable electronic description of the added row model of
the reconstructed surface. In both reduced and reconstructed cases,
pure DFT methods incorrectly pin the defect states inside and to the
lower part of the conduction band (CB) which indicates a tendency for
strong delocalization.

\begin{figure}[htb]
\epsfig{file=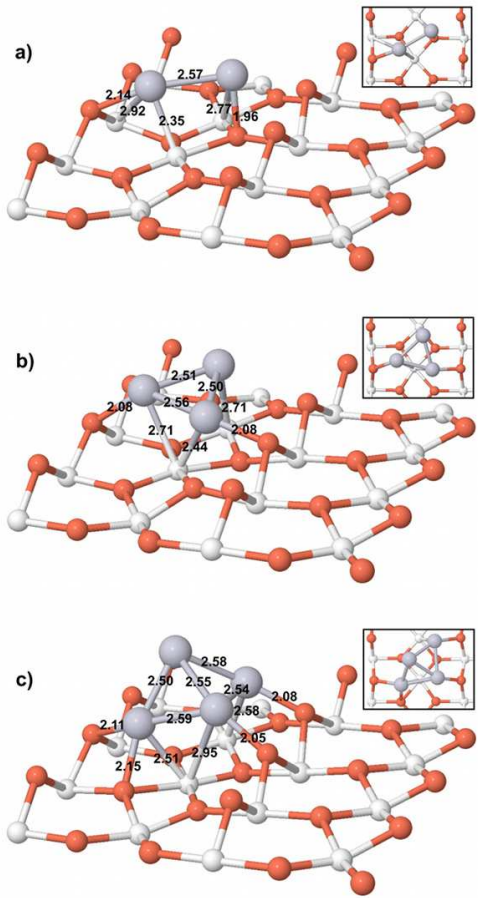,width=7cm} 
\caption{(color online) Minimum energy structures of Pt$_n$
($n$\,=\,2\,--\,4) adsorbed on stoichiometric rutile TiO$_2$(110)
surface. O, Ti, and Pt atoms are denoted by black (red), white
small balls and gray big balls, respectively. Pt dimer, trimer,
and tetramer on this surface are depicted in (a), (b), and (c),
respectively. Measurements are shown, on the corresponding bonds,
all in angstr\"{o}ms. The top view of the small Pt particles are
presented in the insets to provide better description of the
adsorption sites.\label{fig2}}
\end{figure}

\begin{figure}[htb]
\epsfig{file=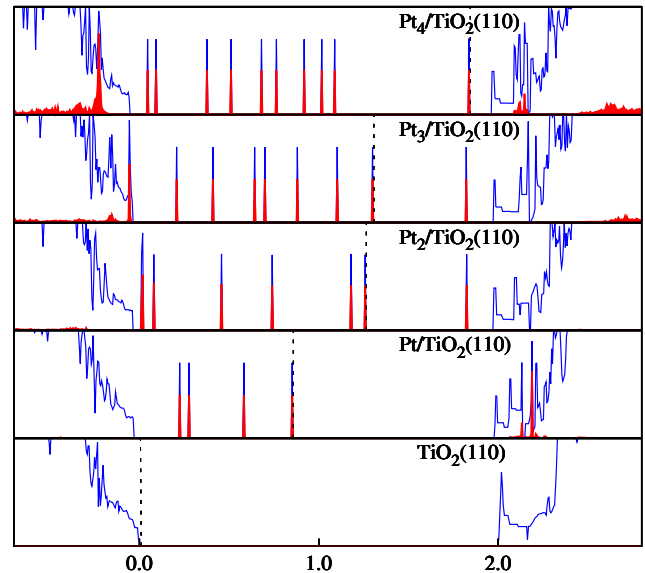,angle=-90,width=9cm,clip=true} 
\caption{(color online) Calculated projected and total density
of states (DOS) structures for Pt$_n$ ($n$\,=\,1\,--\,4) on
stoichiometric rutile TiO$_2$(110)-4$\times$2 surface. Upper
three panels correspond to the atomic structures presented in
Fig.~\ref{fig2}. \label{fig3}}
\end{figure}

GGA+U calculations were performed using 4$\times$2 cell in order to
avoid any charge transfer between the periodic images of Pt clusters
on the bulk terminated (110) surface. Moreover, in order to obtain
the energetics correctly, we first investigated the magnitude of
spin-polarization for small Pt clusters on the surface. The results
gave negligibly small spin multiplicities for the ground states of
TiO$_2$(110) supported Pt$_n$ clusters and for the bare surface,
as well. In particular, similar findings were reported by
Iddir \textit{et al.} for Pt/TiO$_2$(110) system.~\cite{iddir}
Therefore, adsorption profiles [Fig.~\ref{fig2}] and electronic
structures [Fig.~\ref{fig3}] were calculated without spin
polarization for the case of stoichiometric surface.

Relaxed atomic positions for defect free surface are presented
in Table~\ref{table1}. Our GGA and GGA+U values do not considerably
differ from each other and are in good agreement with previous
theoretical calculations~\cite{bates,harrison,sano,kiejna,thompson,hameeuw}
as well as the experimental results.~\cite{charlton,lindsay} The only
discrepancy, which exists for all DFT studies, is seen in Ti6c--B1
bond (B in Fig.~\ref{fig1}). In order to inspect its relevance to the
number of layers, we also performed relaxation calculations with 7
and 8 trilayer cells, that reproduced the same value. Therefore, it
can be well addressed to GGA-PBE exchange--correlation functional that
underestimes Ti6c--B1 in-plane bond length by 0.12 {\AA}. In addition,
experiments reported slightly different results for bond length between
Ti6c and the nearest neighbor subsurface oxygen (C in Fig.~\ref{fig1}).
For this particular bond distance, our GGA and GGA+U values are consistent
with the result of Charlton \textit{et al.}~\cite{charlton} more than
the recent result of Lindsay \textit{et al.}~\cite{lindsay} Agreement
with the experimental data gets better for deeper subsurface bond lengths.


The density of states (DOS) structure for the bare surface is
presented in the first panel of Fig.\ref{fig3}. The Fermi energy is
set to be the zero of energy scale and located just above the valence
band maximum (VBM). The upper part of the VB shows dominant O $2p$
character with a non-negligible Ti $3d$ contribution. Top of the VB
mainly composed of O1 states giving a peak $\sim$0.4 eV below the
VBM. The contribution from the basal oxygens, B1, lies relatively
lower in energy inside the VB. Our site projected DOS analysis for
the VB agrees well with previous studies, (e.g. hybrid PBE0 result
of Labat \textit{et al.}~\cite{labat} and pure GGA result of Sano
\textit{et al.}~\cite{sano}). The lower part of the conduction band
(CB) is dominated by Ti $3d$ states. The bottom of CB shows a hybrid
Ti5c--B1 character with a large Ti contribution. Although, the
composition of the CB have some similarities, we have considerable
disagreement in the shape of the CB edge with other GGA and PBE0
results~\cite{labat,sano}. This is due to on-site $U$ repulsion
acting on the Ti $3d$ electrons that causes charge localization
around the atomic Wigner--Seitz radii. Dudarev's $U$ supplements
correlation energy of $d$ electrons that elevates the corresponding
band offsets to higher energies giving a gap of 2.01 eV. Moreover,
it induces differences in the dispersion of Ti $3d$ states,
particularly, near the conduction band minimum (CBM).

Binding of a single Pt atom to the stoichiometric surface was
investigated for all possible adsorption sites. We found the lowest
energy position of Pt at the hollow site over the Ti5c, leaned
toward the nearest bridging oxygen, O1. This adsorption position was
also reported previously by GGA-PBE calculations~\cite{iddir}. The
calculated BE is 2.32 eV at the hollow site where Pt--O and Pt--Ti
bond lengths become 1.96 and 2.49 {\AA}, respectively. Pt atom pulls
Ti5c slightly up out of the surface plane by 0.15 {\AA}, and also
draws O1 off the bridging oxygen row so that the bond length between
this oxygen and Ti6c extends to 2.04 {\AA} from its bare surface
value of 1.88 {\AA} [in Table~\ref{table1}]. These values are
slightly larger than the GGA-PBE results due to additional on-site
Coulomb repulsion between Ti $3d$ electrons with Dudarev $U=4.5$ eV.
For instance, O1--Ti6c bond length increases by 0.03 {\AA} from its
GGA-PBE value of 1.85 {\AA}.

The DOS for Pt/TiO$_2$(110) presented in Fig.~\ref{fig3} indicates
that single Pt at the hollow site interacts with Ti5c that appears
as a Pt--Ti5c hybridization peak at the bottom of the CB. The
non-bonding excess charge around Pt brings four occupied flat gap
states within a range of 0.9 eV above the VBM. In addition to
the band gap narrowing, these impurity states increases visible
range transition probability upon a vertical excitation. This
will lead to a metal--to--substrate charge transfer, useful for
solar cell applications.

Energetically preferential adsorption geometry of Pt dimer is shown
in Fig.~\ref{fig2}a. This atomic structure happens when a second Pt
is attached to the Pt/TiO$_2$ system as to form a dimer. While one
of the Pt atoms is already at the hollow site the other one finds
its minimum energy position above basal oxygen, B1. It also weakly
pulls the nearest neighbor O1 slightly distorting it out of the
oxygen row at a separation of 2.39 {\AA}. Alternatively, we placed
Pt dimer as a whole on the surface at possible adsorption sites and
found the same geometry between the oxygen rows as the minimum energy
structure. Pt--Pt bond distance becomes 2.57 {\AA} that is meaningfully
larger than the isolated Pt dimer length by 0.24 {\AA}.

Adsorption of Pt trimer breaks its gas phase equilateral symmetry and
causes local deformations at the surface plane and in the second
subsurface layer as seen in Fig~\ref{fig2}b. This geometry is obtained
through Pt$_2$/TiO$_2$ system by an additional Pt atom as to form a
triangular clustering so that it coordinates with the nearest neighbor
O1 at the second oxygen row. Pt$_3$ at different adsorption sites is
energetically not preferable. The two Ti5c under the cluster are lifted
up by $\sim$0.23 {\AA} and B1 atoms, coordinated to these Ti5c, are
pushed down by 0.5 {\AA} from their relaxed clean surface positions.
The longest Pt--Pt bond length occurs as a result of the pulling of the
bridging oxygens from the two sides so that it slightly stretches from
2.50 {\AA} to 2.56 {\AA}. Meanwhile, the shortest Pt--Pt bond is 2.46
{\AA} between the Pt's at the hollow and the Ti5c sites.

Electronically Pt$_2$ and Pt$_3$ on the stoichiometric surface yield
very similar DOS structures except the number and the band energies of
cluster-driven impurity states below the Fermi energy. Pt dimer brings
six occupied states within the range of 1.30 eV above the VB and
an empty flat going gap state 0.15 eV below the CB. Therefore, Pt$_2$ on
the surface causes a band gap narrowing of 1.45 eV with respect to that
of the clean surface. The interaction between the additional Pt at Pt$_3$
and the substrate occurs essentially with the nearest neighbor O1 on
the second oxygen row in Fig.~\ref{fig2}b. It brings two more occupied
flat states, one of them at the VBM, and causes a shift in band energies
of the gap states. In the case of Pt$_3$, metal driven empty gap state
lies almost at the same position with that of Pt$_2$/TiO$_2$(110).
Resulting band gap narrowing is calculated to be 1.50 eV.

Pt--O1 coordination number increases to three as the cluster size adds
up to Pt$_4$. Although it gives the strongest total binding of 2.98 eV
(in Table~\ref{table2}), BE per Pt of 0.75 eV proves to be the lowest
among the other adsorbates. The reason for this is that one of the Pt
atoms lying above the other three makes no contact with the surface.
Pt$_4$ particle on the surface occurs to be slightly distorted
bent-rhombus similar to its isolated GS structure (in Fig.~\ref{fig2}c).
Electronically, Pt$_4$/TiO$_2$(110) system features the lowest energy
band gap by filling the upper lying flat gap state that is 0.13 eV
below the CB. The position of this state is almost the same for the
multi-platinum adsorbates and is empty in the cases of Pt$_2$ and
Pt$_3$ structures. Altogether, Pt$_4$ brings a total of ten occupied
gap states. It couples to the surface by giving significant contribution
to the upper part of the VB essentially by three Pt--O1 hybrid bondings
and at the bottom of the CB by three Pt--Ti5c coordinations.

For all cases, adsorption of small Pt particles causes considerable
local distortions that are mediated by the neighboring atoms to the
second trilayer underneath and are proportional to the cluster size.
Although Pt$_n$ adsorbates induce surface deformations, clusters
themselves seem to be less affected by forming structures similar to
their gas phase low energy geometries on stoichiometric rutile (110)
surface. This indicates that the metal--metal coordination within the
clusters is stronger than the cluster-substrate interaction.

\subsection{Pt$_\mathbf{n}$ on reduced rutile TiO$_\mathbf{2}$(110) surface}

\begin{figure}[htb]
\epsfig{file=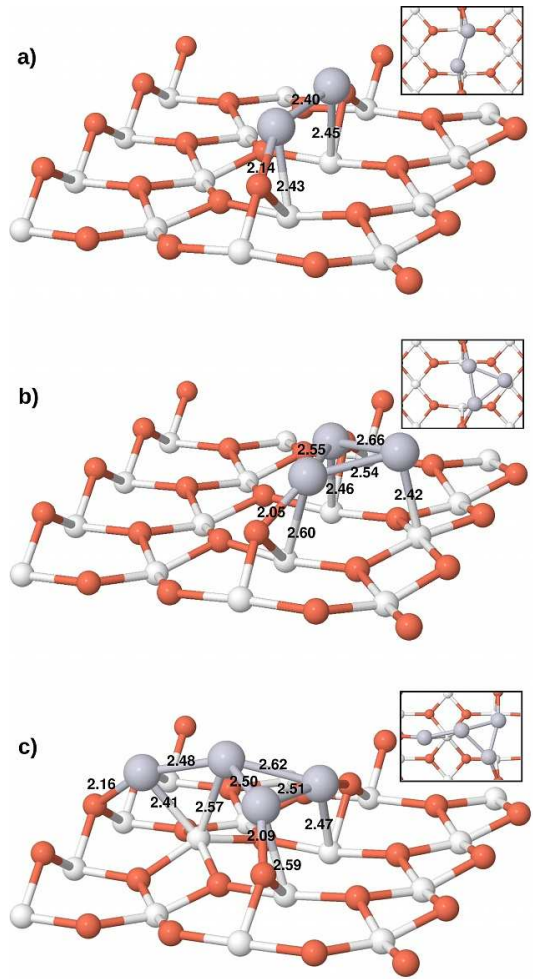,width=7cm}
\caption{(color online) Minimum energy structures of Pt$_n$
($n$\,=\,2\,--\,4) adsorbed on partially reduced rutile
TiO$_2$(110) surface. O, Ti, and Pt atoms are denoted by black
(red), white small balls, and gray big balls, respectively. Pt
dimer, trimer, and tetramer on this surface are depicted in
(a), (b), and (c), respectively. Measurements are shown, on
the corresponding bonds, all in angstr\"{o}ms. The top view of
the small Pt particles are presented in the insets are provided
for visual convenience.\label{fig4}}
\end{figure}
\begin{figure}[htb]
\epsfig{file=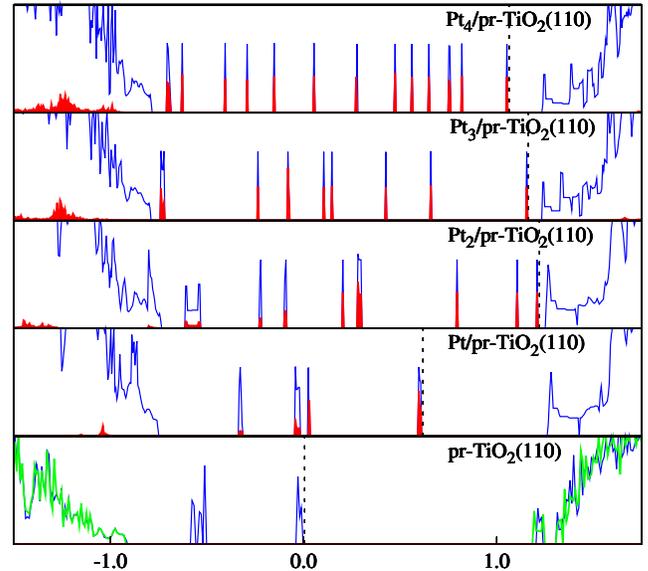,angle=-90,width=9cm,clip=true}
\caption{(color online) Calculated projected and total DOS
structures for Pt$_n$ ($n$\,=\,1\,--\,4) on partially reduced
(pr) rutile TiO$_2$(110)-4$\times$2 surface. Upper three panels
correspond to the adsorption geometries in Fig.~\ref{fig4}.
\label{fig5}}
\end{figure}

Reduced surface is constructed using the relaxed stoichiometric 4$\times$2
cell with 5 trilayers. We removed a single bridging oxygen, O1, from both
surfaces to avoid formation of an unreal dipole across the slab. Isolated
oxygen vacancies are experimentally observed to produce defect states in
the band gap 0.7--0.9 eV below the CB.~\cite{henrich2,gopel,kurtz,henderson}
Pure DFT methods tend to give too delocalized solution where excess charge
density occupies the bottom of the CB. Morgan \textit{et al.}~\cite{morgan}
suggested that a spin polarized GGA calculation, corrected by Hubbard $U$
with $U\geqslant 4.2$ eV reproduces the experimentally observed gap state.
Our choice of $U=4.5$ eV falls within this range and gives the defect state
in the gap $\sim$1.2 eV below the CB as shown in the lower panel of
Fig.~\ref{fig5}. Spin multiplicity at the GS happens to be 4 for the
4$\times$2 cell with two reduced surfaces. The position of the defect states
is sensitive to atomic relaxations that depends on the choice of the Dudarev
$U$ parameter. Calzado \textit{et al.}~\cite{calzado} also showed that gap
states move closer to the CB edge as the oxygen vacancy concentration
increases in agreement with the UPS observations~\cite{henrich2,sadeghi}.

Single Pt atom finds its minimum energy position at the bridging oxygen
vacancy site with equal Pt--Ti6c bond lengths of 2.41 {\AA} each. Pt
oxidizes the surface such that the vacancy induced topographical
deformations recover. The combined system looks like the 4$\times$2
stoichiometric surface cell except that one of the bridging oxygens
replaced by a Pt atom which stays 0.7 {\AA} above the oxygen row level.
Therefore, Pt adsorption at the defect site gives a strong binding of
3.32 eV compared with that of the stoichiometric case.

We also investigated Pt adsorption on fully reduced rutile(110) surface.
At high oxygen vacancy concentrations, an isolated Pt atom takes the place
of the nearest neighbor threefold coordinated basal oxygen, B1; an effect
also known as the strong metal support interaction (SMSI). Then, B1 loses
its coordination with Ti5c and moves to the nearest vacancy site to form
a bridge configuration so that the surface gets oxidized. Pt at B1 site
causes considerable local geometric deformations while B1 becomes a twofold
coordinated O1.

The site projected DOS structure of the Pt/pr-TiO$_2$ system is shown
in Fig.~\ref{fig5}. The contribution of single Pt adsorbate is seen
below the Fermi energy indicated as the dark shade (red). The ground
state of a Pt atom is $d^9s^1$. Therefore, Pt makes two strong bonds
with Ti6c at the vacancy site. The BE is 3.32 eV this is 1 eV larger
than that on the stoichiometric surface. These bonding energy states
fall within the valence band. Pt at the defect site has a significant
influence on the local surface reconstruction. This is clearly seen
when one compares the DOS structures  at the VBM, before and after Pt
adsorption on pr-TiO$_2$. Pt adsorbate also brings four occupied gap
states. Upper lying two of them have strong Pt contribution at 0.78
and 1.37 eV above the VBM. Only the third one below the Fermi energy
shows a weak dispersion. Others are all flat showing non-bonding
character. Lowest lying flat band is 0.43 eV above the VBM.

Adsorption geometries of Pt$_n$ for $n$\,=\,2\,--\,4 on oxygen defective
surface are shown in Fig.~\ref{fig4} where pr-TiO$_2$(110) stands for
partially reduced rutile TiO$_2$(110) surface associated with oxygen
vacancy concentration of 1/8. Pt$_2$ binds to the reduced surface at the
vacancy site as shown in Fig.~\ref{fig4}a.  Each Pt atom relaxes atop
Ti6c site at each side so that the dimer tends to align with the oxygen
row. Pt$_2$ at the defect site pushes Ti6c atoms down from their in
plane positions causing considerable distortion on the Ti6c row. Pt--Pt
bond length shortens to 2.40 {\AA} from its isolated value of 2.52 {\AA}
as a result of strong binding. The BE per Pt atom was calculated to be
1.11 eV that is larger than that of Pt dimer on the stoichiometric
surface due to the oxygen vacancy.

We have traced a number of possible adsorption geometries for Pt$_3$
particle on the defective surface. Pt$_3$/pr-TiO$_2$(110) in
Fig.~\ref{fig4}b represents the minimum energy geometry of the trimer
whose initial structure is obtained by adding a Pt atom to the Pt dimer
in Fig.~\ref{fig4}a. This additional Pt gets adsorbed at Ti5c site and
pulls it up by $\sim$0.40 {\AA} from its in plane position. Resulting
Pt--Ti5c bond length becomes 2.42 {\AA}. Due to imperfect alignment of
the Pt dimer along the surface oxygen row this triangular cluster shape
becomes slightly distorted compared with the equilateral triangle of
gas phase Pt$_3$.

Pt$_2$ and Pt$_3$ adsorbates give very similar DOS characteristics on
the reduced surface. For both cases, eight impurity states fall within
the band gap as shown in Fig.~\ref{fig5}. One expects to get narrower
band gaps as the number of Pt on the surface increases. Interestingly,
Pt$_2$/pr-TiO$_2$(110) system has the narrowest band gap of 0.05 eV
among the other cases. This corresponds to a band gap narrowing
of 1.22 eV relative to the Fermi energy of the pr-TiO$_2$(110).
Although the band gap is underestimated by DFT due to improper
description of the exchange-correlation energy, band gap narrowing
must be absolute and so, experimentally verifiable. This value
has been calculated to be 1.16 eV for the Pt$_3$ case.

In Pt$_4$ case, we examined initial adsorption structures including
low energy gas phase rhombus and pyramid isomers on partially reduced
surface. Relaxed cluster-surface combined system has been found to be
the geometry as shown in Fig.~\ref{fig4}c. Symmetrical rhombus form
can be obtained by adding a Pt atom to Pt$_3$/pr-TiO$_2$ system in
Fig.~\ref{fig4}b. However, after relaxation, such a structure is
energetically less favorable by 0.35 eV/cell. Comparably, Pt$_4$ in
Fig.~\ref{fig4}c, shows an increased coordination with O1's. This
structure can also be obtained by adding a Pt to Pt$_3$/pr-TiO$_2$
so that Pt$_4$ lies between the two bridging oxygen rows. All these
small Pt particles cluster around the vacancy site. Their geometries
are different, in particular for Pt$_4$, from those on the
stoichiometric surface, due to the surface charge distribution around
the defect site. Pt particles oxidize and therefore bind to defective
surface stronger than to the defect free one.

Pt$_4$ adsorption on the reduced surface induces thirteen flat-like
occupied gap states due to the excess charge brought by the metal
cluster, which is larger than that of the previous cases. Fermi
energy occurs at just above the upmost lying flat state as seen
in the top panel of Fig.~\ref{fig5}. This corresponds to a narrowing
of 1.06 eV relative to the band gap of the reduced surface. Since all
Pt clusters bind so as to center around the defect site, all cases 
have contributions to the DOS'es from these bonding Pt--Ti6c
states that appear inside the VB with increasing Pt content. Indeed,
Pt$_4$/pr-TiO$_2$(110) system has the largest partial DOS contribution
inside the VB which essentially disperses similar to that of the Pt$_3$
case.

\begin{table}[htb]
\caption{Calculated total binding energies (eV) of Pt clusters
and corresponding average binding energies per Pt atom at
TiO$_2$(110)-4$\times$2 surfaces.\label{table2}}
\begin{ruledtabular}
\begin{tabular}{lccc}
& stoichiometric & reduced\footnote{with oxygen vacancy concentration $\theta$=0.125} &
reconstructed\footnote{added-row model of Onishi and Iwasawa~\cite{onishi}} \\[1mm] \hline
Pt & 2.32 / 2.32 & 3.32 / 3.32 & 3.94 / 3.94 \\
Pt$_2$ & 1.74 / 0.87 & 2.22 / 1.11 & 4.20 / 2.10 \\
Pt$_3$ & 2.66 / 0.89 & 3.26 / 1.09 & 4.76 / 1.59 \\
Pt$_4$ & 2.98 / 0.75 & 3.38 / 0.85 & 4.54 / 1.14 \\
\end{tabular}
\end{ruledtabular}
\end{table}

\subsection{Pt$_\mathbf{n}$ on reconstructed rutile TiO$_\mathbf{2}$(110) surface}

\begin{figure}[htb]
\epsfig{file=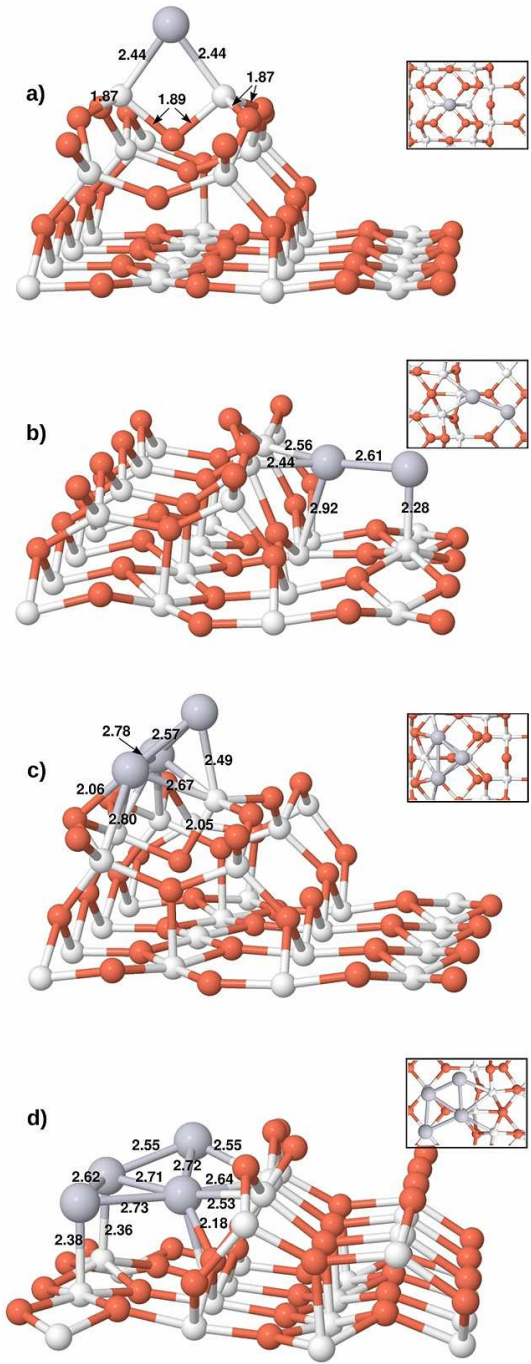,width=7cm,clip=true}
\caption{(color online) Minimum energy structures of reconstructed
rutile TiO$_2$(110) surfaces with Pt$_n$ ($n$\,=\,1\,--\,4) adsorbates.
O, Ti, and Pt atoms are denoted by black (red), white small balls,
and gray big balls, respectively. Bond lengths are shown in angstr\"{o}ms.
The top view of the small Pt particles are presented in the insets are
provided for visual convenience.\label{fig6}}
\end{figure}

The atomic resolution of the long range ($1\times 2$) phase of
reconstructed rutile TiO$_2$(110) surface is difficult with
experiments, numerous theoretical studies attempted to resolve
the controversy regarding the surface morphology~\cite{pang,
elliot1,elliot2}. Scanning tunneling microscopy (STM) and
low-energy electron diffraction (LEED) experiments showed the
existence of Ti$_2$O$_3$ added rows in agreement with the model
proposed by Onishi and Iwasawa~\cite{onishi,pang,mccarty,blancorey1,
blancorey2}. We constructed this model of ($1\times 2$) surface
by forming additional Ti$_2$O$_3$ rows along [001] at the top and
the bottom facets of the relaxed stoichiometric slab with 7 trilayers.
This turns out to be a symmetrical cell which prevents a fictitious
dipole formation across the slab. We studied Pt cluster adsorption
on $4\times 2$ cell that accommodates separations of $\sim$ 12 {\AA}
along [001] and $\sim$ 13 {\AA} along [010] to avoid any charge
transfer between the periodic images of the metal adsorbates. The
relaxed geometries of Pt$_n$/ar-TiO$_2$(110) systems are depicted
in Fig.~\ref{fig6}.

The electronic structure of added row model of the reconstructed
surface was given by Kimura \textit{et al.}~\cite{kimura} and
recently by Blanco-Rey \textit{et al.}~\cite{blancorey1}. By
employing pure DFT methods they obtained DOS peaks resulting from
Ti$_2$O$_3$ row occupying the bottom of the CB. The origin of these
DOS contributions are similar to Ti$^{+3}$ states derived from
oxygen vacancies on the reduced surface. Pure DFT is known to
fail in describing these vacancy states predicting a metallic
character.~\cite{morgan,calzado,nolan} However, defect states
appear 0.7--0.9 eV below the CB.~\cite{henrich2,gopel,kurtz,henderson}
Therefore, for (1$\times$2) reconstruction, pure DFT must incorrectly
predict Ti $3d$ states coming from the Ti$_2$O$_3$ row to appear
inside, at the bottom of, the CB. We tried to obtain a reasonable
electronic description of the added row model of reconstructed,
bare and Pt$_n$ ($n$\,=\,1\,--\,4) adsorbed surfaces by employing
Dudarev $U$ corrected spin polarized DFT calculations. Our tests
showed that calculated band structure of TiO$_2$ is sensitive to
relaxation of the atomic positions based on energy minimization of
the computational cell. In this sense, supplementary $U$ repulsion
energy considerably changes the Ti--O bond lengths, resulting surface
geometry, and so, the band structure.

\begin{figure}[htb]
\epsfig{file=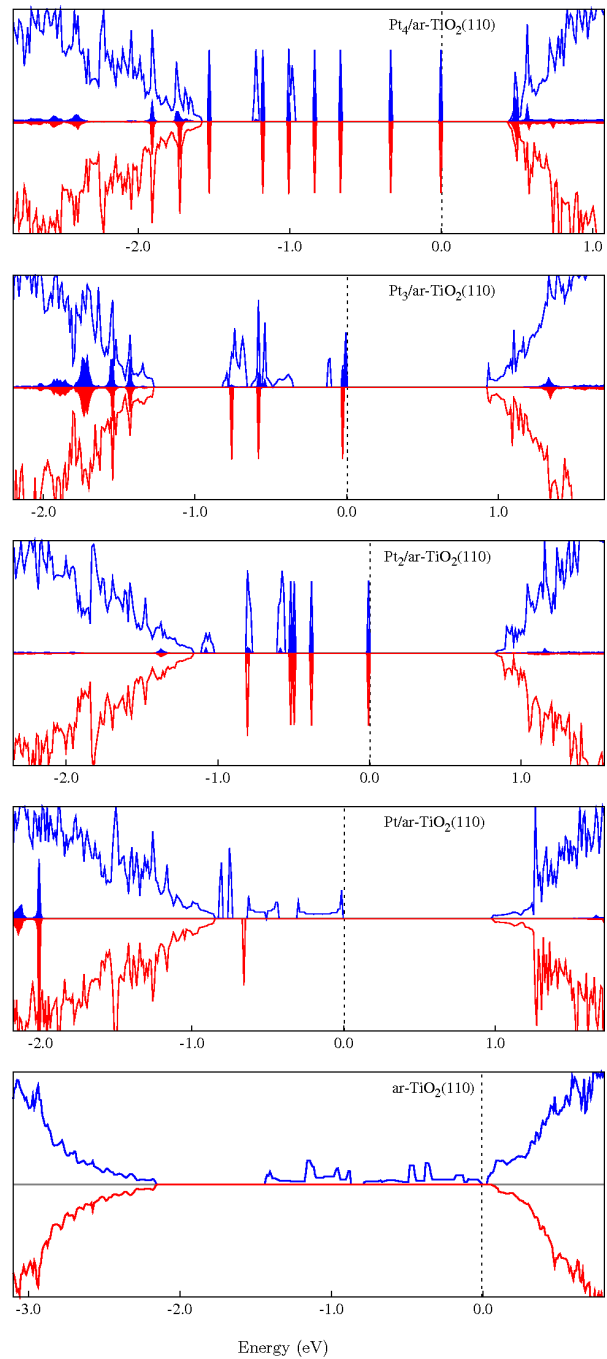,width=7.9cm,clip=true}
\caption{(color online) Calculated projected and total DOS structures for
Pt$_n$ ($n$\,=\,1\,--\,4) on added row (ar) model of reconstructed rutile
TiO$_2$(110)-4$\times$2 surface for the low energy adsorption models shown
in Fig.~\ref{fig6}.\label{fig7}}
\end{figure}

Experimentally, Abad \textit{et al.} observed the band gap state
at 0.7 eV having Ti $3d$ character from the UPS He-I spectra for
TiO$_2$(110)-(1$\times$2)~\cite{abad}. In our calculations, the
choice of $U=4.5$ drives the Ti defect states from the CB edge down
into the gap giving band gap of $\sim$0.1 eV in the bottom panel of
Fig.~\ref{fig7}. Certainly, this value is underestimated due to
the improper cancellation of the self interaction, a well known
artifact of DFT. In addition to the appearance of defect states
due to broken stoichiometry, reconstruction also changes the
dispersion of states at the band edges. Our GGA+U calculations
resulted in a ground state with spin multiplicity of 1.99 per
(1$\times$2) surface cell similar to the case of pr-TiO$_2$(110).
Two satellite DOS peaks appear in the band gap associated with
the excess charge on the Ti$_2$O$_3$ row. These defect states
strongly disperse over a width of $\sim$1.3 eV in the gap and
possess Ti $3d$ character. Fermi level occurs 2.16 eV above the
VBM. If we compare band gap narrowing of 2.16 eV with experimental
gap, we expect (1$\times$2) phase of TiO$_2$(110) to exhibit a
photoemission at an energy consistent with the observation of
Abad \textit{et al.}~\cite{abad}


We considered various possible sites for Pt adsorption on
the reconstructed (1$\times$2 reduced) surface. The minimum
energy surface structure of single Pt adsorption is depicted
in Fig.~\ref{fig6}a. An isolated Pt atom gets adsorbed at the
oxygen site over the terrace in between the two Ti cations
lying along [010] that are also labeled as Ti(iv)~\cite{elliot1}.
Pt pulls these Ti cations, and so, indirectly the oxygen
underneath, forming a symmetrical tetragonal group. The
oxygen elevates up by 1.5 {\AA} from its relaxed Ti$_2$O$_3$
row position. Pt--Ti equilateral bond length becomes 2.44 {\AA}
and Ti--O bonds are 1.89 {\AA} belonging to the tetragon while
the two Ti(iv)'s make four equidistant bonds with oxygens
over the row reading 1.87 {\AA} each. Pt on reconstructed
surface has the strongest binding energy of 3.94 eV/atom
among the other system considered in this study.

Electronically, adsorbate driven atomic dislocations on the
TiO$_2$(110) surface has a significant effect on its band
structure. In particular, band edges are sensitive to any
distortion on the Ti$_2$O$_3$ construction. For example,
single Pt atom adsorption changes the valence and conduction
band edges and offsets of the reconstructed rutile surface
when we compare the corresponding DOS structures of
ar-TiO$_2$(110) systems with and without Pt atom in
Fig.~\ref{fig7}. For an isolated Pt, Fermi energy occurs
at 0.85 eV above the VBM due to upper lying gap state.
The band gap is widened by 0.92 eV in comparison to that
of the clean reconstructed (1$\times$2) surface. Single Pt
atom on 4$\times$2 cell corresponds to a low concentration
coverage. It can only partially reoxidize the Ti(iv) defect
states, which appear in the bottom panel of Fig.~\ref{fig7}
for the majority-spin component. The states associated to
these saturated dangling bonds as a result of Pt binding
fall into the VB at around $-2.0$ eV.  Remaining partially
unsaturated defect states move down to lower energies
leading to the gap widening. Their DOS exhibits two sharp
van Hove singularities with dispersions weaker than those
of the clean reconstructed surface. Non-bonding excess
charge leads to one and two flat-like filled states just
above the VBM for majority and minority spin components,
respectively.

A Pt dimer energetically prefers to get adsorbed at the
trench in interaction with two reactive Ti(iv) cations on
the added row and with the in-plane Ti5c as shown in
Fig.~\ref{fig6}b. In comparison, the total energy of a
4$\times$2 cell with Pt dimer over Ti5c row along [100]
is interestingly just 0.04 eV higher than the lowest energy
geometry. Evidently, Ti5c is an active site inside the
trench. Other cases appear to be far less probable. For
example, Pt$_2$ over and along the Ti$_2$O$_3$ row is
energetically less favorable by 0.92 eV. In Fig.~\ref{fig6}b,
the dimer shows an extended length of 2.61 {\AA} which
is considerably larger than its isolated value of 2.33 {\AA},
essentially due to strong Pt--Ti coupling. As a result of the
Pt--Ti(iv) interactions having 2.44 and 2.56 {\AA} bonds,
the oxygen over the row changes its outward posture and
shows an inward alignment. Moreover, Pt at Ti5c site pulls
the Ti underneath out of its in-plane position by $\sim$0.8
{\AA} leading to a bond length of 2.28 {\AA}. These
dislocations distort the Ti$_2$O$_3$ group breaking
its symmetry along [010].

Pt$_2$ particle brings six majority and five minority flat-like
impurity states below the Fermi energy as shown in Fig.~\ref{fig7}.
In addition, a slightly more dispersing Pt driven state appear
just above the VBM for the spin-up component. Fermi energy
is determined by the upper lying impurity state at 1.17 eV
above the VBM. Thus, the band gap is larger by 0.75 eV relative
to that of ar-TiO$_2$(110). Evidently, the appearance of partial
DOS contribution from mainly 5$d$ electrons of the Pt cluster
in the vicinity of the Fermi energy indicates that Pt$_2$
has non-bonding localized excess charge density which was
absent in the case of an isolated Pt case.

We have investigated various probable adsorption structures
for Pt$_3$ particle. Energetically, Pt trimer favors the
on-the-row adsorption over an in-trench position as shown in
Fig.~\ref{fig6}c. Pt trimer forms in an equilateral triangle
with a 2.78 {\AA} and two 2.57 {\AA} bonds. Pt$_3$ and Pt$_1$
cases have some common features. Topmost Pt atom pulls Ti(iv)
atom so that the bond length becomes 2.49 {\AA} similar to
that of Pt/ar-TiO$_2$(110). The sub-oxygen of the row that is
denoted by its bond with Ti(iv) as 2.05 {\AA} in Fig.~\ref{fig6}c
is pulled up as in the single Pt case. At the other side of the
row Pt--Pt bond aligns with [100] pushing the two oxygens away
so that, that side of the row bends outwards.

The DOS structure of Pt$_3$/ar-TiO$_2$(110) system displays some
similarities with that of Pt$_2$ in that a number of impurity
states appear below the Fermi energy exhibiting dominant Pt 5$d$
character. Evidently, these weakly dispersing impurity states are
brought by the excess charge density localized on the metal
cluster. Similar to the case Pt$_2$, Fermi level is determined
by the upper lying flat-like state at 1.27 eV above the VBM that
corresponds to a band gap widening of 0.90 eV with respect to
bare reconstructed surface. On the other hand the offsets of the
VB edges differ due to adsorbate induced different local surface
relaxations. Pt$_3$ and a single Pt atom stick to the Ti$_2$O$_3$
group. Therefore the partial DOS contribution of bonding 5$d$
electrons appear in the corresponding VBs. Since, the coordination
number of Pt$_3$ with the added row is relatively larger compared
to the Pt case, and since Pt$_3$ distorts the row more than a single
Pt do, bonding Pt$_3$--Ti$_2$O$_3$ states appear in the upper part
of the VB.

The minimum energy structure for the Pt$_4$ case has been found
to be as shown in Fig.\ref{fig6}d where the trench accommodates
the cluster. We see that Pt$_4$ adsorption builds up on the
Pt$_2$/ar-TiO$_2$ skeleton by additional two platinum atoms so
that a gas phase Pt$_4$ forms in the trench. Inward position of
the row oxygen closer to the cluster is similar to that of the
Pt$_2$ case. Two adjacent vertical Pt--Ti5c bonds become 2.38 and
2.36 {\AA} that are slightly larger than the value of 2.28 {\AA}
for the Pt dimer on the surface. They reduce the interaction
strength by relatively better saturating the dangling bonds.
Clearly, off-planar form of the rhombus gets distorted due to
Pt--Ti(iv) interactions so that the side bonds are considerably
extended. As in the case of Pt$_2$ adsorption the added row
construction gets pulled towards Pt$_4$ through Pt-Ti(iv)
interaction. Similar arguments can be made for Pt$_3$ and
Pt--surface cases.

In the top panel of Fig.~\ref{fig7}, the CB edge is determined
by empty Pt$_4$ 6$s$ states. Eight and seven flat-like occupied
impurity states appear in the DOS for the spin up and down
components, respectively, associated mainly with the 5$d$
electrons of Pt$_4$ cluster. The flat impurity state lying
1.59 eV above the VBM sets the Fermi level corresponding
to a band gap larger by 0.28 eV relative to that of the bare
reconstructed surface. The positioning of the cluster--surface
bonding states inside the VB is similar to Pt$_2$ case which
shows common adsorption characteristics.

Pt adsorbates prefer to bind to Ti atoms on the added-row model of
the reconstructed surface since Ti$_2$O$_3$ stoichiometry leads to
oxygen deficiency. Hence, oxygens on the added-row become less
reactive compared to Ti cations. As a trend, odd and even numbered
cluster sizes show different adsorption characteristics. While
clusters with odd number of Pt atoms energetically prefer to be
on the added row terrace, even numbered ones relax into the trench
at one side of the row. Furthermore, it is interesting to see that
Pt$_4$--surface system has common elements with Pt$_2$ case as a
building block, although they were considered distinctly and their
initial structures were not relevant in this sense. Pt clusters in
the trench breaks the symmetrical formation of the added row
construction along [010] while on-the-row adsorption more or less
resumes the symmetry.

\section{CONCLUSIONS}
A systematic analysis of atomic and electronic structure of
small Pt particles, from monomers to tetramers, supported on
regular and defect sites of TiO$_2$(110) surface has been
presented based on Hubbard U corrected hybrid DFT calculations
using Dudarev's approach. The atomic structure of the titania
supported Pt clusters determined by the surface stoichiometry
that constrains the charge density. The calculated binding
energies of Pt particles on the Onishi and Iwasawa model of
the (1$\times$2) surface happens to be greater than those of
the corresponding adsorption systems on partially reduced
(oxygen vacant) and stoichiometric surfaces due to electron
delocalization from the reduced Ti sites. Interestingly, these
chemisorbed Pt clusters form geometries similar to their gas
phase structures except Pt$_4$ at the defect site on the
partially reduced surface where bridging O--Pt interaction
plays a role. In fact, binding energies per Pt atom decreases
as the Pt cluster size increases due to stronger meta-metal
coordination. Thus, large Pt particles show 3D-like nucleation.

We demonstrated that DFT+U method reproduces experimentally observed
gap states for the non-stoichiometric surfaces where electronic
correlation is of vital importance. In particular, Ti(iv) 3$d$ states
have been shown to fall within the band gap of TiO$_2$(110)-(1$\times$2)
reconstructed surface. Those were incorrectly predicted by pure DFT to
be inside the CB giving a metallic character, similar to oxygen
vacancy states. Interaction of small Pt clusters with the TiO$_2$(110)
delocalizes charge causing distortion in the geometry around the
adsorption site. This significantly alters the band edges of the
titania support and also brings band-gap states depending on the
cluster size. The position of these defect states show strong
dependence on the lattice relaxations. Consequently, adsorbate--substrate
coupling leads to significant band gap narrowing for bulk terminated
and partially reduced surfaces while it gives rise to a gap widening
in the case of reconstructed surface. No metallization occurs
for Pt$_n$/TiO$_2$ ($n$\,=\,1\,--\,4) systems. These results provide
good insight into the effect of deposition of small Pt particles on
the atomic and electronic structure of TiO$_2$(110) surfaces with
adsorbates. The detailed analysis of the differences in DOS upon Pt
cluster adsorption puts forward a sound physical picture for
adsorbate-substrate interface properties. In Pt$_n$--TiO$_2$(110)
chemisorption systems, the appearance of well localized non-bonding
Pt impurity states might be useful for catalytic and photovoltaic
applications.

\begin{acknowledgments}
We acknowledge partial financial support from T\"{U}B\.{I}TAK, The
Scientific and Technological Research Council of Turkey (Grant no:
TBAG 107T560). In conjunction with this project, computational
resources were provided by ULAKB\.{I}M, Turkish Academic Network \&
Information Center. V\c{C} and EM also acknowledge secondary support
from Bal{\i}kesir University through BAP project number 2010/37.
\end{acknowledgments}


\begin{thebibliography}{98}
\bibitem{henrich1} V. E. Henrich and P. A. Cox, \emph{The Surface Science of Metal Oxides},
(Cambridge Univ. Press, Cambridge, 1994).

\bibitem{fujishima} A. Fujishima and K. Honda, Nature (London) \textbf{238}, 37 (1972).
\bibitem{diebold1} U. Diebold, Surf. Sci. Rep. \textbf{48}, 53 (2003).
\bibitem{hangfeldt} A. Hangfeldt and  M. Gr\"{a}tzel, Chem. Rev. \textbf{95}, 49 (1995).
\bibitem{gratzel} M. Gr\"{a}tzel, Nature (London) \textbf{414}, 338 (2001).
\bibitem{khan} S. Khan, J. M. Al-Shahry, and W. B. Ingler, Science \textbf{297}, 2243 (2002).
\bibitem{chen}  M. Chen, Y. Cai, Z. Yan, and D. W. Goodman, J. Am. Chem. Soc. \textbf{128}, 6341 (2006).
\bibitem{finetti}  P. Finetti, F. Sedona, G. A. Rizzi, U. Mick, F. Sutara, M. Svec, V. Matolin,
K. Schierbaum, and G. Granozzi, J. Phys. Chem. C \textbf{111}, 869 (2007).
\bibitem{wu} J. M. Wu and C. J. Chen, J. Am. Ceram. Soc. \textbf{73}, 420 (1990).
\bibitem{griffin} G. L. Griffin and K. L. Sieferring, J. Electrochem. Soc. \textbf{137}, 1206 (1990).

\bibitem{see} A. K. See, M. Thayer, and R. A. Bartynski, Phys. Rev. B \textbf{41}, 13722 (1993).
\bibitem{zzhang}  Z. Zhang, S.-P. Jeng, and V. E. Henrich, Phys. Rev. B \textbf{43}, 12004 (1991).
\bibitem{novak} D. Novak, E. Garfunkel, and T. Gustafsson, Phys. Rev. B \textbf{50}, 5000 (1994).
\bibitem{diebold2} U. Diebold, J. F. Anderson, K. O. Na, and D. Vanderbilt,
Phys. Rev. Lett. \textbf{77}, 1322 (1996).
\bibitem{charlton} G. Charlton, P. B. Hoowes, C. L. Nicklin, P. Steadman, J. S. G. Taylor,
C. A. Muryn, S. P. Harte, J. Mercer, R. McGrath, D. Norman, T. S. Turner, and G. Thornton,
Phys. Rev. Lett. \textbf{78}, 495 (1997).
\bibitem{asari} E. Asari, T. Suzuki, H. Kawanowa, J. Ahn, W. Hayami, T. Aizawa, and R. Souda,
Phys. Rev. B \textbf{61}, 5679 (2000).
\bibitem{lindsay} R. Lindsay, A. Wander, A. Ernst, B. Montanari, G. Thornton, and
N. M. Harrison, Phys. Rev. Lett. \textbf{94}, 246102 (2005).

\bibitem{ramamoorthy1} M. Ramamoorthy, R. D. King-Smith, and D. Vanderbilt,
Phys. Rev. B \textbf{49}, 7709 (1994).
\bibitem{ramamoorthy2} M. Ramamoorthy, D. Vanderbilt, and R. D. King-Smith, Phys. Rev.
B \textbf{49}, 16721 (1994).
\bibitem{lindan} P. J. D. Lindan, N. M. Harrison, M. J. Gillan, and J. A. White,
Phys. Rev. B \textbf{55}, 15919 (1997).
\bibitem{kimura} S. Kimura and M. Tsukada, Appl. Surf. Sci. \textbf{130--132}, 587 (1998).
\bibitem{harrison} N. M. Harrison, X. G. Wang, J. Muscat, and M. Scheffler,
Faraday Discuss. \textbf{114}, 305 (1999).
\bibitem{vogtenhuber} D. Vogtenhuber, R. Podloucky, A. Neckel, S. G. Steinemann, and A. J. Freeman,
Phys. Rev. B \textbf{49}, 2099 (1994).
\bibitem{reinhardt} P. Reinhardt and B. A. He{\ss}, Phys. Rev. B \textbf{50}, 12015 (1994).
\bibitem{bates} S. P. Bates, G. Kresse, and M. J. Gillan, Surf. Sci. \textbf{385}, 386 (1997).
\bibitem{evarestov} R. A. Evarestov and A. V. Bandura, Int. J. Quantum Chem. \textbf{96}, 282 (2004).
\bibitem{bandura} A. V. Bandura, D. G. Sykes, V. Shapovalov, T. N. Troung, J. D. Kubicki,
and R. A. Evarestov, J. Phys. Chem. B \textbf{108}, 7844 (2004).
\bibitem{bredow} T. Bredow, L. Giordano, F. Cinquini, and G. Pacchioni, Phys. Rev. B \textbf{70}, 035419 (2004).
\bibitem{sano} H. Sano, G. Mizutani, W. Wolf, and R. Podlucky, Phys. Rev. B \textbf{70}, 125411, (2004).
\bibitem{yfzhang} Y.-F. Zhang, W. Lin, Y. Li, K.-I. Ding, and J.-Q. Li, J. Phys. Chem. B \textbf{109}, 19270 (2005).
\bibitem{kiejna} A. Kiejna, T. Pabisiak, and S. W. Gao, J. Phys.: Condens. Matter \textbf{18}, 4207 (2006).
\bibitem{thompson} S. J. Thompson and S. P. Lewis, Phys. Rev. B \textbf{73}, 073403 (2006).
\bibitem{hameeuw} K. J. Hameeuw, G. Cantele, D. Ninno, F. Trani, and G. Iadonisi, J. Chem. Phys. \textbf{124}, 024708 (2006).
\bibitem{kowalski} P. M. Kowalski, B. Meyer, and D. Marx, Phys. Rev. B \textbf{79}, 115410 (2009).
\bibitem{labat} F. Labat, P. Baranek, and C. Adamo, J. Chem. Theory Comput. \textbf{4}, 341 (2008).

\bibitem{henrich2} V. E. Henrich, G. Dresselhaus, and H. J. Zeiger,
Phys. Rev. Lett. \textbf{36}, 1335 (1976).
\bibitem{gopel} W. G\"{o}pel, J. A. Anderson, D. Frankel, M. Jaehnig, K. Phillips, J.
A. Sch\"{a}fer, and G. Rocker, Surf. Sci. \textbf{139}, 333 (1984).
\bibitem{kurtz} R. L. Kurtz, R. Stockbauer, T. E. Madey, E. Rom\'{a}n, and J. L. De
Segovia, Surf. Sci. \textbf{218}, 178 (1989).
\bibitem{henderson} M. A. Henderson, Surf. Sci. \textbf{400}, 203 (1998).
\bibitem{wendt} S. Wendt, P. T. Sprunger, E. Lira, G. K. H. Madsen, Z. Li, J. O. Hansen, J. Matthiesen,
A. Blekinge-Rasmussen, E. L\ae{}gsgaard, B. Hammer, and F. Besenbacher, Science \textbf{320}, 1755 (2008).
\bibitem{kruger} P. Kruger, S. Bourgeois, B. Domenichini, H. Magnan, D. Chandesris,
P. Le Fevre, A. M. Flank, J. Jupille, L. Floreano, A. Cossaro, A. Verdini, and A. Morgante,
Phys. Rev. Lett. \textbf{100}, 055501 (2008).
\bibitem{kimmel} Greg A. Kimmel and Nikolay G. Petrik, Phys. Rev. Lett. \textbf{100}, 196100 (2008).

\bibitem{morgan} B. J. Morgan and G. W. Watson, Surf. Sci. \textbf{601}, 5034 (2007).
\bibitem{calzado} C. J. Calzado, N. C. Hernandez, and J. F. Sanz, Phys. Rev. B \textbf{77}, 045118 (2008).
\bibitem{nolan} M. Nolan, S. D. Elliot, J. S. Mulley, R. A. Bennett, M. Basham, and P. Mulheran,
Phys. Rev. B \textbf{77}, 235424 (2008).

\bibitem{onishi} H. Onishi and Y. Iwasawa, Surf. Sci. Lett. \textbf{313}, 783 (1994).
\bibitem{guo} Q. Guo, I. Cocks, and E. M. Williams, Phys. Rev. Lett. \textbf{77}, 3851 (1996).
\bibitem{pang} C. L. Pang, S. A. Haycock, H. Raza, P. W. Murray, G. Thornton,
O. G\"{u}lseren, R. James, and D. W. Bullett, Phys. Rev. B \textbf{58}, 1586 (1998).
\bibitem{elliot1} S. D. Elliott and S. P. Bates, Phys. Rev. B \textbf{65}, 245415 (2002).
\bibitem{elliot2} S. D. Elliott and S. P. Bates, Phys. Rev. B \textbf{67}, 035421 (2003).
\bibitem{mccarty}  K. F. McCarty and N. C. Bartelt, Phys. Rev. Lett. \textbf{90}, 046104 (2003).
\bibitem{blancorey1} M. Blanco-Rey, J. Abad, J. M\'{e}ndez, M. F. L\'{o}pez,
J. A. Mart\'{i}n-Gago, and P. L. de Andr\'{e}s, Phys. Rev. Lett. \textbf{96}, 055502 (2006).
\bibitem{blancorey2} M. Blanco-Rey, J. Abad, C. Rogero, J. M\'{e}ndez, M. F. L\'{o}pez,
E. Rom\'{a}n, J. A. Mart\'{i}n-Gago, and P. L. de Andr\'{e}s, Phys. Rev. B \textbf{75}, 081402(R) (2007).
\bibitem{park} K. T. Park, M. Pan, V. Meunier, and E. W. Plummer, Phys. Rev. B \textbf{75}, 245415 (2007).
\bibitem{shibata} N. Shibata, A. Goto, S.-Y. Choi, T. Mizoguchi, S. D. Findlay,
T. Yamamoto, and Y. Ikuhara, Science \textbf{322}, 570 (2008).

\bibitem{gai} Y. Gai, J. Li, S.-S. Li, J.-B. Xia, and S.-H. Wei, Phys. Rev. Lett. \textbf{102}, 036402 (2009).
\bibitem{linsebigler} A. L. Linsebigler, G. Q. Lu, and J. T. Yates, Chem. Rev. \textbf{95}, 735 (1995).
\bibitem{iddir} H. Iddir, S. \"{O}\u{g}\"{u}t, N. D. Browning, and M. M. Disko,
Phys. Rev. B \textbf{72}, 081407(R) (2005); Phys. Rev. B \textbf{73}, 039902(E) (2006).
\bibitem{pillay} D. Pillay and G. S. Hwang, J. Mol. Struct.:THEOCHEM \textbf{771}, 129 (2006).
\bibitem{gong} X.-Q. Gong, A. Selloni, O. Dulub, P. Jacobson, and U. Diebold,
J. Am. Chem. Soc. \textbf{130}, 130 (2008).

\bibitem{vasp} G. Kresse and J. Hafner, Phys. Rev. B, \textbf{47}, 558 (1993).
\bibitem{pbe} J. P. Perdew, K. Burke, and M. Ernzerhof, Phys. Rev. Lett. \textbf{77}, 3865 (1996).
\bibitem{paw1} P. E. Bl\"{o}chl, Phys. Rev. B \textbf{50}, 17953 (1994).
\bibitem{paw2} G. Kresse and D. Joubert, Phys. Rev. B \textbf{59}, 1758 (1999).
\bibitem{dudarev} S. L. Dudarev, G. A. Botton, S. Y. Savrasov, C. J. Humphreys,
and A. P. Sutton, Phys. Rev. B \textbf{57}, 1505 (1998).
\bibitem{mp} H. Monkhorst and J. Pack, Phys. Rev. B \textbf{13}, 5188 (1976).

\bibitem{huda} M. N. Huda, M. K. Niranjan, B. R. Sahu, and L. Kleinman,
Phys. Rev. A \textbf{73}, 053201 (2006).
\bibitem{airola} M. B. Airola and M. D. Morse, J. Chem. Phys. \textbf{116}, 1313 (2002).
\bibitem{taylor} S. Taylor, G. W. Lemire, Y. M. Hamrick, Z. Fu, and M. D. Morse, J.
Chem. Phys. \textbf{89}, 5517 (1988).
\bibitem{grushow} A. Grushow and K. M. Ervin, J. Chem. Phys. \textbf{106}, 9580 (1997).
\bibitem{xiao} Li Xiao and Lichang Wang, J. Phys. Chem. A \textbf{108}, 8605 (2004).
\bibitem{miedema} A. R. Miedema, Z. Metallkd. \textbf{69}, 287 (1978).
\bibitem{yang} S. H. Yang, D. A. Drabold, J. B. Adams, P. Ordejon, and K. Glassford,
J. Phys.: Condens. Matter \textbf{9}, L39 (1997).

\bibitem{sadeghi} H. R. Sadeghi and V. E. Henrich, J. Catal. \textbf{109}, 1 (1988).

\bibitem{abad} J. Abad, C. Rogero, J. Mendez, M. F. Lopez, J. A. Martin-Gago, and E. Rom\'{a}n,
Appl. Surf. Sci. \textbf{234}, 497 (2004).

\end{thebibliography}
\end{document}